\documentclass[aps,prl,preprint]{revtex4}
\usepackage{graphicx}% Include figure files

\renewcommand{\r}{\vec{r}}
  \newcommand{\R}{\vec{R}}
\renewcommand{\P}{{\bf{P}}}

\renewcommand{\k}{\vec{k}}
  
\newcommand{\ket}[1]{\left|#1\right>}
\newcommand{\bra}[1]{\left<#1\right|}
\newcommand{\bk}[2]{\left<#1|#2\right>}
\newcommand{\bak}[3]{\left<#1|#2|#3\right>}

\renewcommand{\exp}[1]{{\textrm{ exp}}\left(#1\right)}
\newcommand{\mathplus}{+}

\renewcommand{\sin}[1]{\textrm{ sin}\left(#1\right)}
\renewcommand{\cos}[1]{\textrm{ cos}\left(#1\right)}

\begin{document}

\title{Unification of Perdew-Zunger Self-Interaction
Correction, DFT+U, and Rung 3.5 Density Functionals}
\author{Benjamin G. Janesko}
\affiliation{Department of Chemistry \& Biochemistry, Texas Christian University, 2800 S. University Dr, Fort Worth, TX 7629, USA} 
 \email{b.janesko@tcu.edu}
\date{\today}

\begin{abstract} 
We unify the Perdew-Zunger self-interaction correction (PZSIC) to approximate density functional theory (DFT), the Hubbard correction DFT+U, and Rung 3.5 functionals within the Adiabatic Projection formalism. We modify the Kohn-Sham reference system, introducing electron self-interaction in selected states.  Choosing those states as localized orbitals, localized atomic states, or states at each point in space recovers PZSIC, DFT+U, and Rung 3.5.  Typical Hubbard U parameters approximate scaled-down PZSIC. A Rung 3.5 variant of DFT+U opens a band gap in the homogeneous electron gas. 
\end{abstract} 

\maketitle 

Kohn-Sham density functional theory (DFT) is arguably the most widely-used
electronic structure approximation across physics, chemistry, and materials
science. DFT models a system of interacting electrons in external potential
$v_{ext}(\r)$, using a reference system of noninteracting Fermions corrected
by a mean-field (Hartree) electron-electron repulsion and a formally exact
exchange-correlation (XC) density functional. 
The ground-state energy $E$ becomes a unique and variational functional of
the electron density $\rho$. Semilocal XC approximations give
\begin{eqnarray}
\label{eq:dft}
E^{DFT}[\rho]&=&T_s[\rho]+\int
d^3\r\rho(\r)v_{ext}(\r)+U[\rho]+E^{SL}_{XC}[\rho_\alpha,\rho_\beta] . 
\end{eqnarray}
 
Self-interaction (delocalization) error is a major limitation of semilocal XC
approximations. This error leads to incorrect nonzero electron-electron
interaction energies for one-electron systems, and over-stabilization of open
systems of fluctuating electron number.\cite{MoriSanchez2008,Perdew2017} 
Self-interaction correction can improve predictions of band gaps, reaction
barriers, excited states, electron affinities, and other properties. However,
self-interaction error often mimics aspects of electron correlation, and
correcting it can degrade predictions for lattice constants, bond energies, and
strongly correlated systems.\cite{Perdew2009,Zope2019} 
Strategies for addressing this paradoxical\cite{Zope2019}
zero-sum\cite{Janesko2017} tradeoff include the Perdew-Zunger self interaction
correction (PZSIC),\cite{Perdew1981} Hubbard model DFT+U
methods,\cite{Anisimov1993,Dudarev1998,Cococcioni2005} and Rung 3.5
functionals.\cite{Janesko2021}
 
The PZSIC ensures an approximate XC functional returns zero
electron-electron interaction energy in any one-electron system: 
\begin{eqnarray}
\label{eq:sic}
E[\rho](PZSIC) &=& E^{DFT}[\rho] + \sum_{i=1}^N U[\rho_i]+E_{XC}^{SL}[\rho_i,0] . 
\end{eqnarray}
One-electron densities $\rho_{i\sigma}=|\phi_{i\sigma}|^2,\
\sigma=\alpha,\beta$ obey $\rho_\sigma=\sum_i\rho_{i\sigma}$. States
$\phi_{i\sigma}=\sum_jL_{ij}^\sigma\psi_{j\sigma}$ are chosen as a localizing unitary
transform ${\bf{L^\sigma}}$ of the occupied spin-orbitals of the Kohn-Sham
reference system $\psi_{i\sigma}$.\cite{Pederson1984}   $U[\rho]=\frac{1}{2}\int d^3\r_1\int
d^3\r_2\rho(\r_1)|\r_1-\r_2|^{-1}\rho(\r_2)$ denotes the Hartree electron repulsion
of density $\rho$. 
 
DFT+U adds a Hubbard-type model atop a DFT calculation, penalizing fractional
({\em{i.e.}}, delocalized) occupancy of a set of predefined atomic states
$\{\phi_m\}$.  We consider the simplified rotationally invariant scheme:\cite{Dudarev1998,Cococcioni2005}
\begin{eqnarray}
\label{eq:DFTU}
E[\rho](DFT+U) &=& E^{DFT}[\rho] + \sum_{m=1}^M \frac{U_m}{2} \left(n_{m\sigma}-n_{m\sigma}^2\right) . 
\end{eqnarray}
Here $n_{m\sigma}=\sum_i|\bk{\phi_m}{\psi_{i\sigma}}|^2$ is the occupancy of
the $m$th atomic state. (We assume throughout that atomic states are
orthogonalized and occupation matrices are diagonal.) Hubbard parameter
$U_m$ controls the energetic penalty applied to state $m$. Various schemes for
determining $U_m$ have been proposed.\cite{Cococcioni2005,Agapito2015}
 
Our Rung 3.5 functionals project the Kohn-Sham one-particle density matrix
$\gamma_{\sigma}(\r,\r')=\sum_i\psi_{i\sigma}(\r)\psi_{i\sigma}^*(\r')$  onto
localized states $\phi(\r-\r_g)$ centered at each point in space $\r_g$, and
use the projection  as an ingredient in nonlocal approximate XC
functionals: 
\begin{eqnarray}
\label{eq:R35}
n_{\sigma}(\r_g) &=& \int d^3\r\gamma_\sigma(\r_g,\r)\phi_{R35}(\r_g-\r). 
\end{eqnarray}
Global, local, and range-separated hybrid functionals tune the tradeoffs of
self-interaction by incorpoating a fraction of localizing,
one-electron-self-interaction-free exact exchange.\cite{Csonka2010} 

Today, global and screened hybrid functionals are widely adopted and necessarily
empirical,\cite{Csonka2010} DFT+U is widely adopted in solid-state
physics,\cite{Kulik2015} local hybrids and PZSIC are experiencing renewed
interest,\cite{Zope2019} and Rung 3.5 functionals are under active
development.\cite{Verma2019}  Formally, hybrid functionals can be derived
through a generalized adiabatic connection,\cite{Savin1988,Becke1993} Rung 3.5
functionals introduce an upper bound within this derivation,\cite{Janesko2012}
and the PZSIC and DFT+U are based on physical arguments and model Hamiltonians. 
DFT+U is connected to hybrid functionals and some correlated
methods.\cite{Jiang2009,Ivady2014,Gani2016} 
Additional formal connections between these approaches could drive new
developments.

The present work presents a unified derivation of PZSIC, DFT+U, and Rung 3.5
methods, in terms of a reference system experiencing only electron
self-interaction.  The derivation uses the Adiabatic Projection
formalism\cite{Janesko2022a} employed in our generalization of
PZSIC.\cite{Janesko2022b} 
We begin with the general case. Let $\{\ket{\phi_{m\sigma}}\}$ be a set of
normalized $\sigma$-spin one-electron states. For each state, introduce a
weight $w_{m\sigma}$ and define a two-electron projection
$\hat{P}^{2}_{m\sigma}=\ket{\phi_{m\sigma}\phi_{m\sigma}}\bra{\phi_{m\sigma}\phi_{m\sigma}}$
where
$\bk{\r_1\r_2}{\phi_{m\sigma}\phi_{m\sigma}}=\phi_{m\sigma}(\r_1)\phi_{m\sigma}(\r_2)$. 
Consider a system of $N=\sum_{\sigma}N_\sigma$ electrons in external potential
$v_{ext}(\r)$, with Hamiltonian $\hat{H}=\hat{T} +\hat{V}_{ee}+\sum_{i=1}^N
v_{ext}(r_i)$ and electron-electron interaction operator
$\hat{V}_{ee}=\frac{1}{2}\sum_{i,j=1}^N|\r_i-\r_j|^{-1}$. Define a projected
electron-electron interaction operator and a generalized Hamiltonian
\begin{eqnarray}
\label{eq:VP} 
\hat{V}^P_{ee} &=& \sum_{m\sigma} w_{m\sigma} \hat{P}^2_{m\sigma}\hat{V}_{ee}\hat{P}^2_{m\sigma}, \\ 
\hat{H}[\lambda,\P] &=& \hat{T} + \sum_{i=1}^N v_{ext}(\r_i) + \hat{V}_{ee}^P + \lambda\left(\hat{V}_{ee}-\hat{V}_{ee}^P\right) . 
\end{eqnarray}
Here $\P=\{\phi_{m\sigma},w_{m\sigma}\}$ denotes the states and weights, 
$\lambda=0$ corresponds to the reference system, and $\lambda=1$ corresponds to
the real system.  Define $\Psi[\lambda,\P,\rho]$ as the $N$-electron
wavefunction that minimizes the expectation value of $\hat{H}[\lambda,\P]$
while returning density $\rho$. The Hohenberg-Kohn theorems ensure that there
exists some unique and variational density functional 
which ensures that the reference system's ground-state energy and density equal
those of the real system, 
\begin{eqnarray}
\label{eq:HXC}
U[\P,\rho]+E_{XC}[\P,\rho] &=& %
\int_0^1d\lambda\left<\Psi[\lambda,\P,\rho]\right|\hat{V}_{ee}-\hat{V}^P_{ee}\left|\Psi[\lambda,\P,\rho]\right> .  
\end{eqnarray}
We assume the necessary N-representability conditions: for any $\lambda$, we
assume that the ground-state density of the real system can be obtained from a
pure ground state of $\hat{H}[\lambda,\P]+v_\lambda$ for some one-electron
potential $v_\lambda$. 
 
The reference system's
electron-electron interaction energy is zero, its ground-state wavefunction is
a single Slater determinant, and the expectation value of
$\hat{H}[\lambda=0,\P]$ equals that of 
%is the $T_s[\rho]+\int d^3\r\rho(\r)v_{ext}(\r)$ of
standard DFT. Consider $N$-electron Slater
determinant $\ket{\Phi}$ formed from spin-orbitals $\{\psi_{i\sigma}\}$.
The projection $\bk{\phi_{m\sigma}\phi_{m\sigma}}{\Phi}$ becomes
$\sum_{ij}\left(c_{im\sigma}c_{jm\sigma}-c_{jm\sigma}c_{im\sigma}\right)$, where
$c_{im\sigma}=\bk{\phi_{m\sigma}}{\psi_{i\sigma}}$. 
The expectation value of the $m$th term in Eq \ref{eq:VP} becomes 
\begin{eqnarray}
\label{eq:PhiP}
\bak{\Phi}{\hat{P}^2_{m\sigma}\hat{V}_{ee}\hat{P}^2_{m\sigma}}{\Phi} &=& \sum_{i,j=1}^{N\sigma} c^*_{im\sigma}c^*_{jm\sigma}U[\rho_{m\sigma}]\left(c_{im\sigma}c_{jm\sigma}-c_{jm\sigma}c_{im\sigma}\right), \\ 
&=& n^2_{m\sigma} U[\rho_{m\sigma}] - n^2_{m\sigma}U[\rho_{m\sigma}]. \nonumber 
\end{eqnarray}
Here 
$n_{m\sigma}=\sum_i|c_{im\sigma}|^2$ is the occupancy of the $m$th projection
state and 
$U[\rho_{m\sigma}]=\bak{\phi_{m\sigma}\phi_{m\sigma}}{\hat{V}_{ee}}{\phi_{m\sigma}\phi_{m\sigma}}$
is its Hartree self-repulsion.  The self-Hartree and self-exchange terms cancel
to ensure eq \ref{eq:PhiP} is zero.
The natural definition for the projected Hartree term in eq \ref{eq:HXC} is the
difference between the Hartree energies of reference and real systems,
$U[\P,\rho]=U[\rho]-\sum_{m\sigma}w_{m\sigma}n^2_{m\sigma}U[\rho_{m\sigma}]$. 
The ground-state energy of the real system becomes 
\begin{eqnarray}
\label{eq:main}
E[\rho] &=& T_s[\rho] + \int d^3\r v_{ext}(\r) \rho(\r)  + U[\rho] 
-\sum_{m\sigma} w_{m\sigma} n_{m\sigma}^2U[\rho_{m\sigma}] + E_{XC}[\P,\rho] . 
\end{eqnarray}

Everything up to now is exact. To proceed, we choose the states and weights
$\P$ and an approximation for the projected XC functional $E_{XC}[\P,\rho]$.
 
{\em{PZSIC}}: We choose $N_\sigma$ orthonormal localized projection states
$\ket{\phi_{m\sigma}}=\sum_{i}L_{mi}^{\sigma}\ket{\psi_{i\sigma}}$ matching the localized
states in eq \ref{eq:sic}. This choice ensures that all $n_{m\sigma}=1$. 
We approximate  $E_{XC}[\P,\rho]$ by combining states and weights $\P$ with an
existing semilocal XC functional, 
\begin{eqnarray}
\label{eq:xcsic}
E_{XC}[{\bf{P}},\rho](PZSIC) &=& E^{SL}_{XC}[\rho_\alpha,\rho_\beta] -\sum _{m\sigma}w_{m\sigma} E_{XC}^{SL}[\rho_{m\sigma},0] . 
\end{eqnarray}
Choosing weights $w_{m\sigma}=1$ and substituting eq \ref{eq:xcsic} and
$n_{m\sigma}=1$ into eq \ref{eq:main} recovers the original PZSIC of eq
\ref{eq:sic}. Choosing $w_{m\sigma}=1/2$ recovers the scaled SIC of Kl\"upfel
and coworkers.\cite{Kluepfel2012} Choosing $w_{m\sigma}=\int d^3\r\
(\tau_{W\sigma}(\r)/\tau_\sigma(r)) \rho_{m\sigma}(\r)$ recovers the scaled SIC
of Vydrov and Scuseria.\cite{Vydrov2006} (Here
$\tau_\sigma(\r)=\sum_i|\nabla\psi_{i\sigma}(\r)|^2$ and
$\tau_{W\sigma}(\r)=|\nabla\rho_\sigma(\r)|/(4\rho_\sigma(\r))$. )

\begin{figure}
\includegraphics[width=0.9\textwidth]{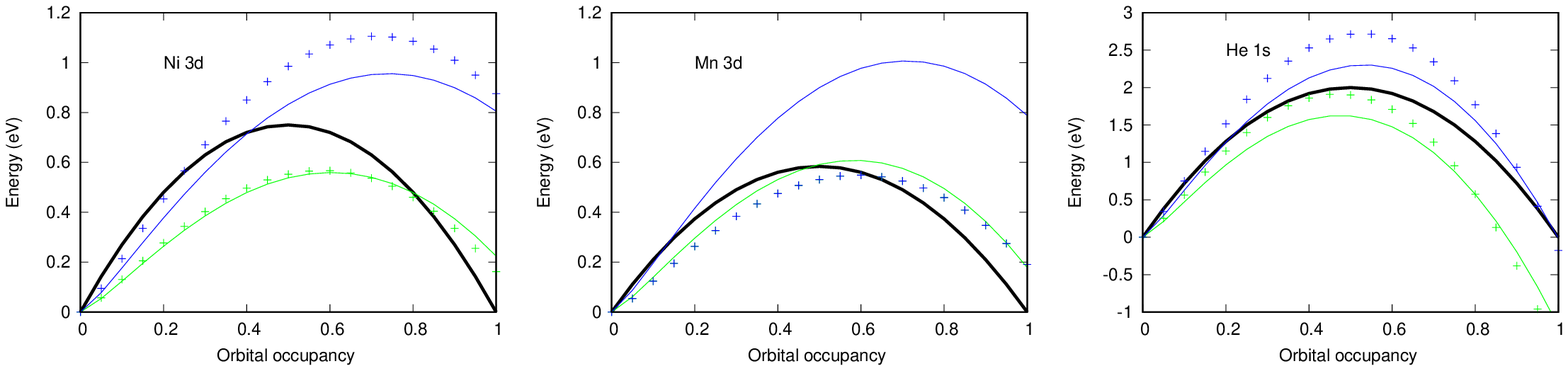}
\caption{\label{fig:U} DFT+U corrections (black thick line), PZSIC-LDA
corrections (green), and PZSIC-PBE corrections (blue) for individual orbitals
(eV), plotted as functions of orbital occupancy. 
(Left) 25\%
rescaled PZSIC for Ni (lines) and Ni$^{2+}$ (points) 3d orbitals, and $U$=6 eV.  
(Center) 25\%
rescaled PZSIC for Mn$^{4+}$ (lines) and Mn$^{2+}$ (points) 3d orbitals, and $U$=4.67 eV.  
(Right)
Unscaled PZSIC for He (lines) and He$^+$ (points) 1s orbital, and $U=16$ eV. }
\end{figure} 

{\em{DFT+U}}: We choose projection states $\ket{\phi_{m\sigma}}$ as localized and
orthogonalized atom-centered states. For example, in a calculation on solid
nickel oxide NiO, we would choose five projection states of each spin per unit
cell, corresponding to the five nickel $3d$ orbitals.  
We approximate  $E_{XC}[{\bf{P}},\rho]$ following eq \ref{eq:xcsic},
noting that occupancies $n_{m\sigma}$ are no longer guaranteed to
be 1.  Substitution into eq \ref{eq:main} and rearrangement yields 
\begin{eqnarray}
\label{eq:pzu} 
E[\rho] &=& 
E^{DFT}[\rho]+\sum_{m\sigma}w_{m\sigma}U[\rho_{m\sigma}]\left(  \frac{E_{XC}^{SL}[n_{m\sigma}\rho_{m\sigma},0]}{U[\rho_{m\sigma}]}  - n_{m\sigma}^2\right) 
\end{eqnarray}
To recover eq \ref{eq:DFTU}, we further assume that the XC functional is linear
in occupancy $n_{m\sigma}$ and perfectly cancels self-interaction at integer
occupancy ($E_{XC}^{SL}[n_{m\sigma}\rho_{m\sigma},0]\simeq
n_{m\sigma}U[\rho_{m\sigma}]$), and define the Hubbard parameter as
$U_m=2w_{m\sigma}U[\rho_{m\sigma}]$. 
 
With this derivation in hand, the magnitude of the Hubbard $U_m$
parameter for atomic state $\ket{\phi_{m\sigma}}$ depends on
the state's unscreened self-Coulomb interaction $U[\rho_{m\sigma}]$, weight
$w_{m\sigma}$, and linearized semilocal approximate XC self-interaction.  Our
result is distinct from previous rationalizations of $U_m$, which invoke
screening or renormalization to account for the effects of the other electrons
in the system.\cite{Ivady2014,Agapito2015,Shih2012}  
Figure \ref{fig:U} directly compares DFT+U to PZSIC, evaluated for fractional
state occupancies using eq \ref{eq:pzu}. PZSIC calculations use HF/def2-TZVP
orbitals of isolated integer-charge high-spin atoms, computed with the PySCF
package.\cite{Weigend2005,Sun2017}
The $U_{eff}=5-6$ eV employed in DFT+U simulations of nickel
oxide\cite{Cococcioni2005,Bengone2000} approximately corresponds to a
scaled-down PZSIC ($w_{m\sigma}=1/4$) of the nickel 3d orbitals. Gratifyingly,
this is consistent with the 1/4 scaling of exact exchange in the
Heyd-Scuseria-Ernzerhof screened hybrid.\cite{Heyd2003} The $U_{eff}=16$ eV
used to correct fractional charge error of isolated helium atom\cite{Bajaj2017}
approximately corresponds to full PZSIC ($w_{m\sigma}=1$) for the helium 1s
orbital. DFT+U generally lies between the PZSIC computed with different choices
of projection state\cite{Pickett1998} ({\em{e.g.}}, neutral {\em{vs.}} cationic
nickel) and XC functional ({\em{e.g.}} LDA\cite{Vosko1980} {\em{vs}}.
PBE\cite{Perdew1996}). 

{\em{Rung 3.5}}: We choose an infinite number of projection states
$\ket{\phi_\sigma(\r_m)}$, one centered at each point in space $\r_m$.
$\bk{\r}{\phi_\sigma(\r_m)}=\phi_\sigma(\r-\r_m)$ denotes the state's value at
point $\r$.  The sum over states $m$ in eq \ref{eq:main} becomes an integral
over points $\r_m$, the weights $w_\sigma(\r_m)$ have units of (length)$^{-3}$, and
$U[\rho_m]$ becomes $U[|\phi_\sigma(\r-\r_m)|^2]$, the Hartree self-repulsion
of state $\phi_\sigma(\r-\r_m)$. The occupancies $n_{m\sigma}$ in eq \ref{eq:main}
become
Rung 3.5 projections similar to eq \ref{eq:R35}:
\begin{eqnarray}
n_\sigma(\r_m) &=& \sum_i\left|\bk{\psi_{i\sigma}}{\phi_\sigma(\r_m)}\right|^2, \\ 
&=& \int d^3\r_1\int d^3\r_2\gamma_\sigma(\r_1,\r_2)\phi_\sigma(\r_m-\r_1)\phi_\sigma(\r_m-\r_2).  \nonumber 
\end{eqnarray}
Choosing the projection states to be Gaussians
$\phi^G_\sigma(\r-\r_m)=(2\alpha/\pi)^{3/4}\exp{-\alpha|\r-\r_m|^2}$ 
recovers Rung 3.5 methods similar to those in Ref. \citenum{Verma2019}.
(Expanding the KS orbitals in $\{\chi_\mu(\r-\R_\mu)\}$, a set of Gaussian-type
basis functions centered at points $\R_\mu$, gives analytically evaluable\cite{Janesko2014}
$n_\sigma(\r_m)=\sum_{\mu\nu}A_\mu(\r_m)P^{\sigma}_{\mu\nu}A_\nu(\r_m)$ where
$A_\mu(\r_m)=\int d^3\r\chi_\mu(\r-\R_\mu)\phi^G(\r-\r_m)$ and
$\gamma_\sigma(\r,\r')=\sum_{\mu\nu}\chi_\mu(\r)P^\sigma_{\mu\nu}\chi_\nu(\r')$.)
Choosing instead a homogeneous electron gas (HEG) model for the projection states
$\phi^L_\sigma(\r-\r_m)=\rho^{-1/2}_\sigma(\r_m)\gamma^{LDA}(\rho_\sigma(\r_m),|\r-\r_m|)$
ensures that the occupancies $n_\sigma(\r_m)$ are independent of uniform density scaling.
Here $4\pi^2\gamma^{LDA}(\rho,u)=\sin{k_Fu}/u^3-k_F\cos{k_Fu}/u^2$ 
denotes the one-particle spin-density matrix of a HEG with Fermi vector 
$k_F=(6\pi^2\rho_\sigma(\r_g))^{1/3}$.

At this point, one can make make many choices for the Rung 3.5 projected XC
functional.  Here we introduce a ``DFT+R35U'' approach based on eq
\ref{eq:pzu}, 
\begin{eqnarray}
\label{eq:R35U} 
E[\rho](DFT+R35U) &=& E^{DFT}[\rho] + \sum_\sigma \int d^3\r_m u(\r_m)\left(n_{\sigma}(\r_m)-n^2_\sigma(\r_m)\right). 
\end{eqnarray} 
The derivation of eq \ref{eq:R35U} parallels that of eq \ref{eq:pzu}. Energy
density $u(\r_m)$ is analogous to Hubbard energy $U_m$.  This analogue of DFT+U
contains no atom-centered states and no atom-dependent parameters. 

\begin{figure}
\includegraphics[width=0.45\textwidth]{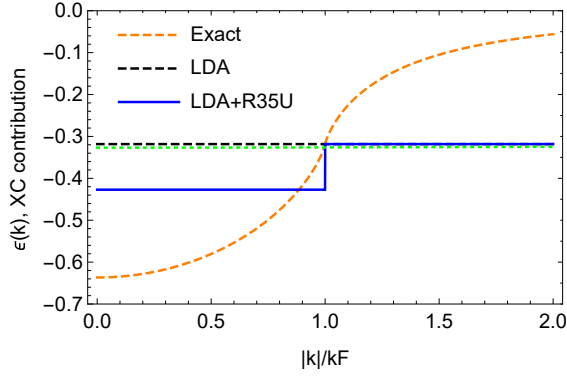}
\caption{\label{fig:R35U} XC contribution to band energy
$\epsilon(\k)$ of the HEG. Exact exchange, LDA exchange, and 
LDA+R35U with Hubbard-like energy density $u=-0.1\rho_\sigma v_{X\sigma}^{LDA}$.  }
\end{figure}

We can apply eq \ref{eq:R35U} to the HEG, a system with $N_\sigma$ electrons in
macroscopic volume $V$ giving translationally invariant density
$\rho_\sigma(\r_m)=\rho_\sigma=N_\sigma/V$
and Kohn-Sham orbitals $\bk{\r}{\k}=V^{-1/2} e^{i\k\cdot\r}$ occupied for
$|\k|\leq k_F$. In contrast to eq \ref{eq:R35}, DFT+U cannot be applied to the HEG because there are no atomic
states, and PZSIC applied to the HEG gives an incorrect total
energy.\cite{Santra2019a}   Choosing constant $u(\r_m)=u$ ensures translational
invariance. 
Choosing projection states $\phi^L_\sigma(\r-\r_m)$ ensures that the projection
onto KS orbitals $\bk{\k}{\phi^L_\sigma(\r_m)}$ is $\rho_\sigma^{-1/2}$  for
occupied states $|\k|<k_F$ and zero for unoccupied states. This ensures that
all projection states are fully occupied ($n_{\sigma}(\r_m)=\int^{kF}d^3\k
|\bk{\k}{\phi^L_\sigma(\r_m)}|^2=1$) and guarantees that eq \ref{eq:R35U}
recovers $E^{DFT}[\rho]$.  The nonlocal potential
$\bak{\r_1}{\hat{v}_{R35U}}{\r_2}$= $\int d^3\r_m
u(\r_m)(1-2n_\sigma(\r_m))\phi_\sigma^L(\r_1-\r_m)\phi_\sigma^L(\r_m-\r_2)$,
defined from 
the functional derivative of eq \ref{eq:R35U} with respect to
$\gamma_\sigma(\r_1,\r_2)$, gives band energy contribution
$\bak{\k}{\hat{v}_{R35U}}{\k}$ equal to $-u\rho^{-1}_\sigma$ for occupied
states and 0 for unoccupied states.  Figure \ref{fig:R35U} illustrates the band
energy contributions from exact exchange, LDA exchange, and LDA+R35U exchange
choosing $u=-0.1 \rho_\sigma v_X^{LDA}$. The R35U correction shifts the
occupied states and opens a bandgap without changing the total energy. Put
another way, DFT+R35U provides an alternative derivation of a ``scissor
operator'' applied to the HEG.\cite{Baraff1984} 

We have generalized the PZSIC using projection onto active spaces, rather than
individual orbitals, to introduce correlation into the reference
system.\cite{Janesko2022b} Here we briefly comment on similar generalizations
of DFT+U and Rung 3.5 methods. 
For DFT+U, we can introduce interactions between the $\alpha$-spin and $\beta$-spin
electrons in atomic state $\phi_m$ using
$P^{2}_m=\sum_{\sigma,\sigma}\ket{\phi_{m\sigma}\phi_{m\sigma'}}\bra{\phi_{m\sigma}\phi_{m\sigma'}}$.
The reference system now contains an opposite-spin interaction
$U[\rho_m]n_{m\alpha}n_{m\beta}$ alongside the self-interaction
$U[\rho_{m\sigma}](n_{m\sigma}^2-n_{m\sigma}^2)$. 
For Rung 3.5 methods, we introduce interactions between $\alpha$-spin and
$\beta$-spin electrons in state $\ket{\phi(\r_m)}$ centered at $\r_m$. The
reference system now contains an integral over the opposite-spin interactions
$\int d^3\r_m u(\r_m)n_\alpha(\r_m)n_\beta(\r_m)$ alongside the
self-interaction. 
In either case, the reference system's electron-electron interaction energy is
no longer zero, and its ground-state wavefunction is generally not a single
Slater determinant. Approximating the reference wavefunction as a single Slater
determinant yields analogues of the opposite-spin terms in DFT+U+J and
``judiciously modified DFT'' generalizations,\cite{Liechtenstein1995,Bajaj2019}
and the opposite-spin Rung 3.5 correlation in M11plus,\cite{Verma2019} which is based on
Becke's real-space nondynamical correlation model.\cite{Becke2003}

\bibliographystyle{aip}
%\bibliography{../../review/tosubmit/tosubmit}

\end{document}